\documentclass[a4paper,11pt]{article}
\usepackage{pos}
\usepackage{xspace}

\newcommand{\x}{\mathrm{x}}
\newcommand{\y}{\mathrm{y}}
\newcommand{\z}{\mathrm{z}}
\renewcommand{\i}{\mathrm{i}}

\renewcommand{\Im}{\mathrm{Im}}

\newcommand{\dd}{\mathrm{d}}

\newcommand{\bN}{\overline{N}}
\newcommand{\bF}{\overline{F}}

\newcommand{\bA}{\bar{A}}
\newcommand{\bB}{\overline{B}}

\newcommand{\nbrho}{\mathrel{\smash{\overset{\makebox[0pt]{\mbox{\tiny (--)}}}{\varrho}}}}

\newcommand{\nbN}{\mathrel{\smash{\overset{\makebox[0pt]{\mbox{\tiny (---)}}}{N}}}}

\newcommand{\nbvF}{\mathrel{\smash{\overset{\makebox[0pt]{\mbox{\tiny (---)}}}{\vec{F}}}}}

\newcommand{\vrho}{\varrho}

\newcommand{\flash}{\texttt{FLASH}\xspace}
\newcommand{\emu}{\texttt{Emu}\xspace}

\renewcommand{\vec}{\mathbf}

\title{Neutrino flavor transformation with moments: application to fast flavor instabilities in neutron star mergers}
\ShortTitle{Neutrino flavor transformation with moments: application to FFIs in NSMs}

\author*[a,b]{Julien Froustey}
\author[c]{Sherwood Richers}
\author[b]{Evan Grohs}
\author[b]{Samuel D. Flynn}
\author[d]{Francois Foucart}
\author[b]{James P. Kneller}
\author[b]{Gail C. McLaughlin}

\affiliation[a]{Department of Physics, University of California Berkeley,\\
Berkeley, CA 94720, USA}

\affiliation[b]{Department of Physics, North Carolina State University,\\
  Raleigh, NC 27695, USA}

\affiliation[c]{Department of Physics, University of Tennessee Knoxville,\\
Knoxville, TN 37996, USA}

\affiliation[d]{Department of Physics \& Astronomy, University of New Hampshire,\\
9 Library Way, Durham, NH 03824, USA}

\emailAdd{jfroustey@berkeley.edu}

\abstract{Neutrino evolution, of great importance in environments such as neutron star mergers (NSMs) because of their impact on explosive nucleosynthesis, is still poorly understood due to the high complexity and variety of possible flavor conversion mechanisms. In this study, we focus on so-called "fast flavor oscillations", which can occur on timescales of nanoseconds and are connected to the existence of a crossing between the angular distributions of electron (anti)neutrinos. Based on the neutrino radiation field drawn from a three dimensional neutron star merger simulation, we use an extension of the two-moment formalism of neutrino quantum kinetics, and perform a linear stability analysis to determine the characteristics of fast flavor instabilities across the simulation. We compare the results to local (centimeter-scale) three-dimensional two-flavor simulations using either a moment method or a particle-in-cell architecture. We get generally good agreement in the instability growth rate and typical instability lengthscale, although the imperfections of the closure used in moment methods remain to be better understood.}

\FullConference{XVIII International Conference on Topics in Astroparticle and Underground Physics (TAUP2023)\\
 28.08--01.09.2023\\
University of Vienna\\}

\begin{document}
\maketitle

\section{Moment neutrino evolution equations and linear stability analysis}

A statistical ensemble of (anti)neutrinos is commonly described in cosmological and astrophysical environments by (one-body reduced) density matrices $\nbrho \! \! (t, \vec{r}, \vec{p})$, which are Hermitian matrices in flavor space. The diagonal components generalize classical distribution functions, while the off-diagonal components account for flavor coherence. Our goal is to study neutrino evolution using angular moments. The first three moments (number density, flux and pressure) are defined by:
\begin{equation}
\label{eq:moments}
    \begin{pmatrix}
        N_{\alpha \beta} \\
        F_{\alpha \beta}^i \\
        P_{\alpha \beta}^{ij}
    \end{pmatrix}(t,\vec{r}) \equiv \int{\dd p} \, \frac{p^2}{(2 \pi)^3} \int{\dd \Omega} \begin{pmatrix} 1 \\ p^i/p \\ p^i p^j/p^2 \end{pmatrix} \vrho_{\alpha \beta}(t, \vec{r}, \vec{p}) \, .
\end{equation}
The evolution of $\nbrho$ is dictated by the Quantum Kinetic Equations (QKEs)~\cite{SiglRaffelt,Blaschke:2016xxt,Froustey:2020mcq}, which we can rewrite in terms of the angular moments~\eqref{eq:moments}. For this study, we consider mono-energetic neutrinos, neglect collisions and restrict the Hamiltonian governing flavor transformation to the self-interaction mean-field part, which is responsible for fast flavor instabilities (FFIs) when an electron lepton number crossing is present~\cite{Richers_review}. Under these assumptions, the moment QKEs read:
\begin{align}
\i \left(\partial_t N + \partial_j F^j \right) &= \sqrt{2} G_F \left[N - \bN^*, N\right]  - \sqrt{2} G_F  \left[(F-\bF^*)_j,F^j\right] \, ,  \label{eq:QKE_moment_N} \\
\i \left(\partial_t F^i + \partial_j P^{ij} \right) &= \sqrt{2} G_F \left[N - \bN^*, F^i\right] 
- \sqrt{2} G_F \left[(F-\bF^*)_j,P^{ij}\right] \, , \label{eq:QKE_moment_F}
\end{align}
with similar equations for antineutrinos. These are only the first two equations of an infinite hierarchy, that we truncate thanks to a \emph{closure relation} $P^{ij}(N,\vec{F})$. For consistency with the NSM simulation we are studying, we use a semi-classical generalization of the maximum entropy closure (MEC) — see~\cite{Grohs:2023pgq,Froustey:2023skf} for details. In the following, we restrict to two-flavor oscillations between the $e$ flavor and the $x$ flavor (heavy lepton flavor neutrinos).

\paragraph{Linear stability analysis (LSA)} In order to determine if Eqs.~\eqref{eq:QKE_moment_N}--\eqref{eq:QKE_moment_F} lead to a flavor conversion instability, we add flavor off-diagonal sinusoidal perturbations to the moments, such that:
\begin{equation}
    N = \begin{pmatrix}
            N_{ee} & A_{e x}e^{- \i (\Omega t - \vec{k} \cdot \vec{r})} \\
            A_{x e}e^{- \i (\Omega t - \vec{k} \cdot \vec{r})} & N_{x x}
        \end{pmatrix} \ , \qquad 
    F^j = \begin{pmatrix}
            F^j_{ee} & B^j_{e x}e^{- \i (\Omega t - \vec{k} \cdot \vec{r})} \\
            B^j_{x e}e^{- \i (\Omega t - \vec{k} \cdot \vec{r})}& F^j_{xx}
        \end{pmatrix} \, .
\end{equation}
At first order in the perturbations, the linearized equations~\eqref{eq:QKE_moment_N}--\eqref{eq:QKE_moment_F} read $(S_{\vec{k}} + \Omega \mathbb{I}) \cdot Q = 0$, with $Q = (A_{ex},B_{ex}^\x,B_{ex}^\y,B_{ex}^\z,\bA_{xe},\bB_{xe}^\x,\bB_{xe}^\y,\bB_{xe}^\z)$ the vector of perturbations (see details in \cite{Froustey:2023skf}). We call $S_{\vec{k}}$, which depends on the wavevector $\vec{k}$ considered, the “stability matrix”. Non-zero solutions for $Q$ are obtained by numerically solving $\mathrm{det}\left(S_{\vec{k}} + \Omega \mathbb{I}\right) = 0$, which has eight solutions for $\Omega(\vec{k})$, of which the eigenmode with the largest value of $\Im(\Omega)$ will dominate. We scan for all values of $\vec{k}$, such that, for a given set of classical moments, the fastest growing mode is:
\begin{equation}
     \Im(\Omega)_\mathrm{max} \equiv \underset{\vec{k}}{\mathrm{max}} \big\{ \Im\left[\Omega(\vec{k})\right] \big\} \qquad \text{corresponding to a wavevector} \  \vec{k}_\mathrm{max} \, . 
\end{equation}

\section{Search for fast flavor instabilities in a neutron star merger}

We apply our linear stability analysis on the results of the classical general relativistic two-moment radiation hydrodynamics simulation of the merger of two $1.2 \, M_\odot$ neutron stars from~\cite{Foucart:2016rxm}. This simulation provides the set of classical moments $(\nbN_{ee},\nbN_{xx},\nbvF_{ee},\nbvF_{xx})$ for each point in a box of size $(136 \, \mathrm{km} \times 136 \, \mathrm{km} \times 68 \, \mathrm{km})$. We focus on the results from a snapshot taken $5 \, \mathrm{ms}$ post-merger.

\begin{figure}[ht]
    \centering
    \includegraphics[width=0.82\textwidth]{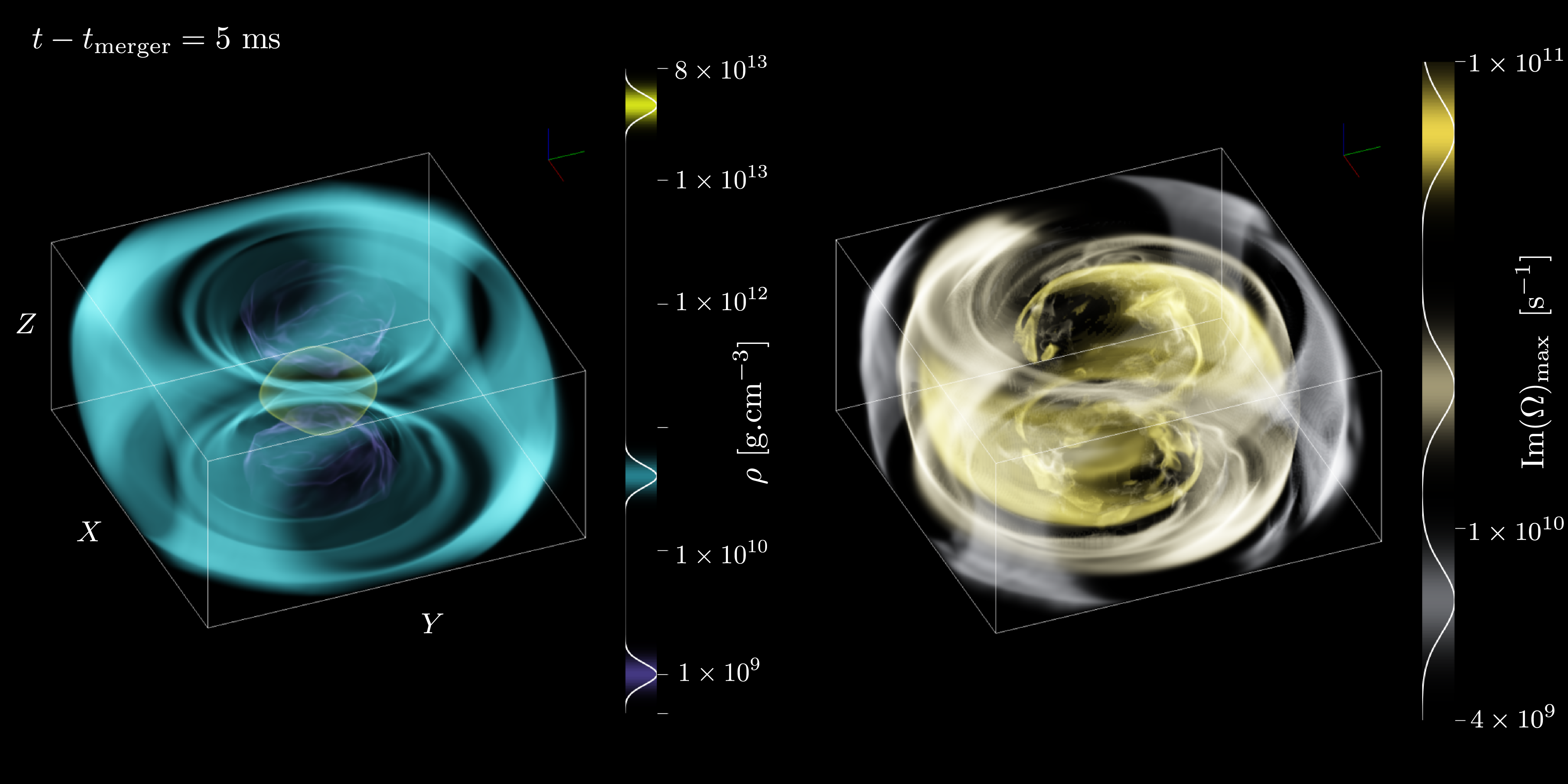}
    \caption{\label{fig:3D_NSM} \emph{Left:} 3D volume rendering of the matter density in the snapshot taken $5 \, \mathrm{ms}$ post-merger in the simulation~\cite{Foucart:2016rxm}. \emph{Right:} volume rendering of the FFI growth rate predicted with moment-LSA. Three colored contours are centered respectively around the growth rate values $7 \times 10^9 \, \mathrm{s^{-1}}$, $2 \times 10^{10} \, \mathrm{s^{-1}}$, and $7 \times 10^{10} \, \mathrm{s^{-1}}$.}
\end{figure}

The snapshot we study and the results from LSA across the simulation are represented Fig.~\ref{fig:3D_NSM}. We identify a “structure” of FFI, with generally smaller growth rates as we go further away from the remnant (which is expected since the neutrino density decreases).

\begin{figure}[ht]
    \centering
    \includegraphics[width=\textwidth]{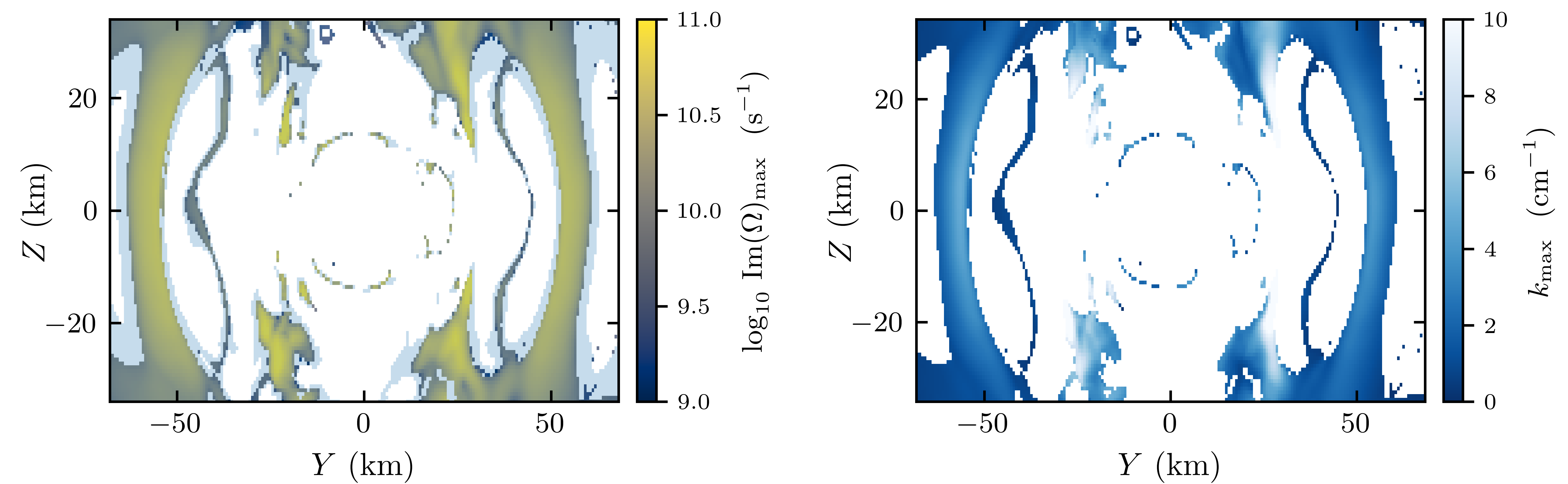}
    \caption{\label{fig:slice} Predictions from moment LSA for the slice $\{X = 6 \, \mathrm{km}\}$. \emph{Left:} instability growth rate and regions where an ELN crossing is predicted (light blue areas). \emph{Right:} wavenumber of the fastest growing mode.}
\end{figure}

We show in Fig.~\ref{fig:slice} the predicted characteristics of FFI on a slice of the simulation, taken at constant $X \simeq 6 \, \mathrm{km}$ (other slices are presented in~\cite{Froustey:2023skf}). These confirm the smallness of time- ($\sim 0.1 \, \mathrm{ns}$) and length- ($\sim \mathrm{cm}$) scales of FFI in a NSM, which shows the challenge an inclusion of FFI in a large-scale hydrodynamics simulation represents. In addition, we compare the unstable regions found from LSA with the known criterion of existence of an electron lepton number (ELN) crossing. The regions where such a crossing exists can be determined for the MEC following~\cite{Johns:2021taz,Richers:2022dqa}. These regions are superimposed in light blue on the left panel of Fig.~\ref{fig:slice}. There is generally an excellent agreement, which is a success since the moment method only has access to a very limited angular information. 
We attribute the differences to the imperfections of our closure~\cite{Froustey:2023skf}.

\paragraph{Quantitative comparison} In order to assess the quality of our LSA, we compare its predictions with full non-linear numerical simulations of the QKEs, based either on a moment method (\flash) or a particle-in-cell multi-angle architecture (\emu). Results are shown in Table~\ref{tab:NSM1} for the “NSM 1” point located at $(X \simeq 20 \, \mathrm{km}, \, Y = 0 \, \mathrm{km}, \, Z \simeq 20 \, \mathrm{km})$ and studied in~\cite{Grohs:2022fyq,Grohs:2023pgq}. The overall agreement is satisfying ; the remaining differences can be attributed to the details of the moment implementation in \flash, and the limitations of the maximum entropy closure that is used throughout moment calculations, but only to set the initial conditions in \emu.

\renewcommand{\arraystretch}{1.1}

\begin{table}[!ht]
    \centering
    \begin{tabular}{|r|c|c|c|}
    \hline
         & LSA & \flash & \emu  \\ \hline \hline
        $\quad \Im(\Omega)_\mathrm{max} \ (10^{10} \ \mathrm{s}^{-1})$ & $7.25$  & $8.1$ & $5.6$ \\
        $k_\mathrm{max} \ (\mathrm{cm}^{-1})$ & $5.68$ & $6.4(4)$ & $4.8(4)$ \\  \hline
    \end{tabular}
    \caption{\label{tab:NSM1} Linear stability analysis~\cite{Froustey:2023skf} and simulation~\cite{Grohs:2022fyq,Grohs:2023pgq} results for the “NSM 1” point of the simulation.}
\end{table}

\section*{Conclusion}

A two-moment method with closure offers a more computationally efficient way to study FFI compared to multi-angle calculations, and LSA allows us to predict the overall characteristics with good accuracy. There are some clear shortcomings, most of them due to the imperfection of the closure relation~\cite{Froustey:2023skf}, but the flexibility and time-efficiency of LSA will allow for a more comprehensive study and assessment of better closure prescriptions in the future.

Using moments is an important improvement of our description of neutrino evolution in neutron star mergers, as it connects directly to the quantities used in classical simulations. Along with its intrinsic reduced computation time, these features open a new path towards the \emph{in situ} inclusion of neutrino flavor transformation in large-scale simulations. To this end, future research will need to tackle the design of a proper \emph{quantum} closure, and include other flavor conversion mechanisms~\cite{Volpe:2023met}, such as “slow” modes, matter-neutrino resonances, collisional instabilities...

\bibliographystyle{JHEP}
\bibliography{references}

\providecommand{\href}[2]{#2}\begingroup\raggedright\begin{thebibliography}{10}

\bibitem{SiglRaffelt}
G.~Sigl and G.~Raffelt, \emph{{General kinetic description of relativistic mixed neutrinos}}, \href{https://doi.org/10.1016/0550-3213(93)90175-O}{\emph{Nucl. Phys. B} {\bfseries 406} (1993) 423}.

\bibitem{Blaschke:2016xxt}
D.N.~Blaschke and V.~Cirigliano, \emph{{Neutrino Quantum Kinetic Equations: The Collision Term}}, \href{https://doi.org/10.1103/PhysRevD.94.033009}{\emph{Phys. Rev. D} {\bfseries 94} (2016) 033009} [\href{https://arxiv.org/abs/1605.09383}{{\ttfamily 1605.09383}}].

\bibitem{Froustey:2020mcq}
J.~Froustey, C.~Pitrou and M.C.~Volpe, \emph{{Neutrino decoupling including flavour oscillations and primordial nucleosynthesis}}, \href{https://doi.org/10.1088/1475-7516/2020/12/015}{\emph{JCAP} {\bfseries 12} (2020) 015} [\href{https://arxiv.org/abs/2008.01074}{{\ttfamily 2008.01074}}].

\bibitem{Richers_review}
S.~Richers and M.~Sen, \emph{{Fast Flavor Transformations}},  in \emph{{Handbook of Nuclear Physics}}, I.~Tanihata, H.~Toki and T.~Kajino, eds., (Singapore), pp.~1--17, Springer Nature Singapore (2022), \href{https://doi.org/10.1007/978-981-15-8818-1_125-1}{DOI} [\href{https://arxiv.org/abs/2207.03561}{{\ttfamily 2207.03561}}].

\bibitem{Grohs:2023pgq}
E.~Grohs, S.~Richers, S.M.~Couch, F.~Foucart, J.~Froustey, J.~Kneller et~al., \emph{{Two-Moment Neutrino Flavor Transformation with applications to the Fast Flavor Instability in Neutron Star Mergers}},  \href{https://arxiv.org/abs/2309.00972}{{\ttfamily 2309.00972}}.

\bibitem{Froustey:2023skf}
J.~Froustey, S.~Richers, E.~Grohs, S.~Flynn, F.~Foucart, J.P.~Kneller et~al., \emph{{Neutrino fast flavor oscillations with moments: linear stability analysis and application to neutron star mergers}},  \href{https://arxiv.org/abs/2311.11968}{{\ttfamily 2311.11968}}.

\bibitem{Foucart:2016rxm}
F.~Foucart, E.~O'Connor, L.~Roberts, L.E.~Kidder, H.P.~Pfeiffer and M.A.~Scheel, \emph{{Impact of an improved neutrino energy estimate on outflows in neutron star merger simulations}}, \href{https://doi.org/10.1103/PhysRevD.94.123016}{\emph{Phys. Rev. D} {\bfseries 94} (2016) 123016} [\href{https://arxiv.org/abs/1607.07450}{{\ttfamily 1607.07450}}].

\bibitem{Johns:2021taz}
L.~Johns and H.~Nagakura, \emph{{Fast flavor instabilities and the search for neutrino angular crossings}}, \href{https://doi.org/10.1103/PhysRevD.103.123012}{\emph{Phys. Rev. D} {\bfseries 103} (2021) 123012} [\href{https://arxiv.org/abs/2104.04106}{{\ttfamily 2104.04106}}].

\bibitem{Richers:2022dqa}
S.~Richers, \emph{{Evaluating approximate flavor instability metrics in neutron star mergers}}, \href{https://doi.org/10.1103/PhysRevD.106.083005}{\emph{Phys. Rev. D} {\bfseries 106} (2022) 083005} [\href{https://arxiv.org/abs/2206.08444}{{\ttfamily 2206.08444}}].

\bibitem{Grohs:2022fyq}
E.~Grohs, S.~Richers, S.M.~Couch, F.~Foucart, J.P.~Kneller and G.C.~McLaughlin, \emph{{Neutrino fast flavor instability in three dimensions for a neutron star merger}}, \href{https://doi.org/10.1016/j.physletb.2023.138210}{\emph{Phys. Lett. B} {\bfseries 846} (2023) 138210} [\href{https://arxiv.org/abs/2207.02214}{{\ttfamily 2207.02214}}].

\bibitem{Volpe:2023met}
M.C.~Volpe, \emph{{Neutrinos from dense environments : Flavor mechanisms, theoretical approaches, observations, and new directions}},  \href{https://arxiv.org/abs/2301.11814}{{\ttfamily 2301.11814}}.

\end{thebibliography}\endgroup

\end{document}